# A NOVEL METHOD FOR FATIGUE TESTING OF MEMS DEVICES CONTAINING MOVABLE ELEMENTS

Z. Szűcs, M. Rencz

*Budapest University of Technology and Economics (BME), Department of Electron Devices*
H-1521 Budapest, Goldmann Gyorgy ter 3., Hungary
Fax: +36 1 463 2973
szucs@eet.bme.hu, rencz@eet.bme.hu

**ABSTRACT**

*In this paper we present a novel method for reliability testing of MEMS devices containing movable structures. A small size, simple and cheap vibration fatigue test equipment was designed and realized at BUTE and vibration fatigue tests were carried out on 10 samples of a LIS0L02AS4-type MEMS 3-axis inertial sensor provided by ST Microelectronics. The paper presents the test plan, the test equipment and the results with a detailed statistical comparison at the end. The work has been carried out within the framework of Patent-DfMM project. At different framework locations the same devices were tested by different methods in order to compare the test equipments and the results.*

## 1. INTRODUCTION

The fact that MEMS structures use not only electrical effects for their operation poses several new reliability issues beyond those well-known problems we can see in integrated circuit technology. Since MEMS can have many moving elements, their reliability and lifetime is also a big concern. Stiction, wear, fatigue, thermal degradation and package defects are the most often encountered problems that can significantly affect the lifetime of these structures [1, 2]. The ultimate benefit of every work in this field is to find better solutions in future design to maintain longer lifetime and higher reliability in Micro-Electromechanical Systems.

Vibration tests are a group of fatigue tests where the load or one of the environmental parameters e.g. temperature is cyclically changing. The aim of these tests is to detect the effect of the cyclical changes on the structure [3]. MEMS fatigue tests are usually time consuming and often destructive tests. They are suitable for the observation of the structure's behavior in different circumstances (e.g. harsh environment or vacuum). Important reliability data, like lifetime or mean time to failure (MTTF) and several Design for Reliability issues can be obtained from them.

Standard vibration fatigue test equipments are usually big and expensive devices designed for PCB and miscellaneous electronic equipment testing. As MEMS devices are small in size, they can be easily vibrated by only a simple loudspeaker. Several kinds of non-standard vibration fatigue test equipments can be found in the literature. These are usually hand-made experimental devices for a specific application [4, 5, 6]. Our unique test arrangement provides fast and easy vibration testing with the possibility of in-situ inspection of MEMS devices.

## 2. THE PATENT-DFMM PROJECT

The work of the Patent-DfMM network aims to establish a collaborative team to provide European industry with support in the field of "Design for Micro and Nano Manufacture" to investigate the problems affecting the manufacture and reliability of products based on micro- and nanotechnologies [7]. Within this network there are several collaborative groups of research teams. One of these is working on comparing and benchmarking different reliability testing methods. All the members of this group investigate the properties of a MEMS inertial sensor, produced and provided by ST Microelectronics. After the separate work of each group the results will be evaluated and compared to each other. Conclusions concerning the applicability of test methods will be made, the advantages and drawbacks of the different methods will be determined.

The device under test is a 3-axis linear accelerometer (LIS3L02AS4) that includes a sensing element and an electronic interface able to take the information from the sensor and provide an analog signal output. The sensing element is a surface micro-machined capacitive half-bridge, and its electronic interface is manufactured using a standard CMOS process [8].

## 3. TEST PLAN

The purpose of the test is to determine the effect of vibration on the device in a specified frequency range. The detailed test circumstances are defined in the related MIL-STD-883F-2005.2 test method [9]. According to the description, the device shall be rigidly fastened on the vibration platform and the leads shall be adequately secured. A constant amplitude harmonic vibration shall be applied for 32 ±8 hours minimum, in each of the orientations X, Y, and Z. The frequency of the vibration should be in the range of 60 ±20 Hz having a peak acceleration of 20 g. The peak acceleration can be calculated by Eq. 1.

$$n_g = \frac{A \cdot \omega^2}{g} = \frac{A \cdot 4\pi^2 f^2}{g} \approx 4.0243 \cdot A \cdot f^2 \quad \text{(Eq. 1)}$$

Where A is the amplitude [m], $\omega$ is the angular velocity [1/s], f is the frequency [1/s] and g≈9.81 m/s$^2$.

For a peak acceleration of 20g (Test condition A) [9] an amplitude of 1 mm was applied at a frequency of 80Hz.

## 4. VIBRATION TESTS

As a first step, basic non-destructive characterization steps were done in order to check the operability of all the 10 samples. Afterwards vibration fatigue tests were carried out at room temperature with in-situ inspection of the samples.

### 4.1. Our test equipment

A specific vibration tester (VIBROTEST) was designed and built at the Department of Electron Devices of BUTE, Hungary. Its main features are the following:
- Computer controlled test
- Frequency range: 1 Hz...40 kHz
- Programmable frequency, amplitude, waveform, duty-cycle
- Able to test 4 samples at a time and measure (life)time in cycles and/or hours
- Optional failure detection

Figure 1 shows the block diagram of the system. The microcontroller unit communicates with the computer via the serial port (COM1-RS232). The user can have the highest level control over the whole system through the PC. The *Function Generator* provides a stable 80 Hz sine wave for the controller. Once the test has been started, the *Power Amplifier* amplifies this signal to a proper level in order to drive the *Resonator* with the *Specimen* mounted on it. The three DC *Power Supply Units* (PSU) provide the necessary power to operate the circuits and the *Oscilloscope* enables us to verify the output waveforms of the device under test.

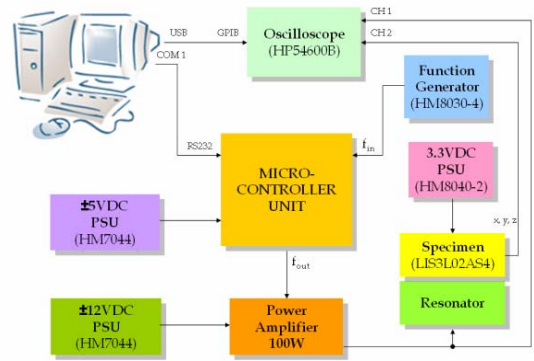

Fig.1. Block diagram of VIBROTEST

In order to clamp the specimen to be tested, we have designed a vibration platform (Fig. 3.). This is a small FR4 PCB with pads. With the help of this, in situ testing of the accelerometer is soluble.

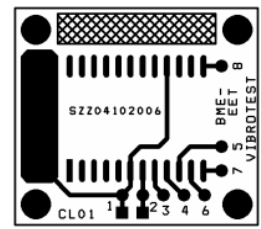

Fig. 3. The vibration platform layout for LIS3L02AS4

If one wants to connect the device electrically for in-situ testing, pads 1 to 8 should be connected to the test circuitry by thin, enamelled copper wires. Fig 4. shows specimen No.1 fastened on the vibration platform in direction X.

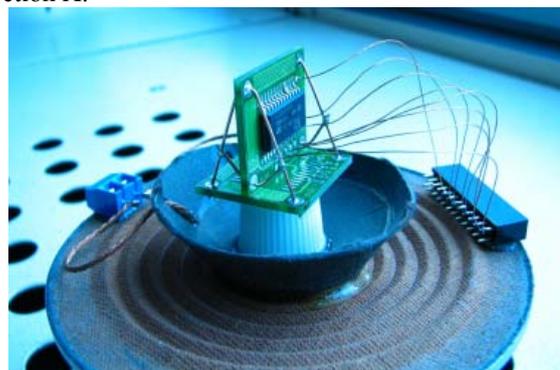

Fig. 4. Specimen No.1 on the vibration platform

**4.2 Test Results**

All of the 10 MEMS accelerometers were tested according the above given standard test specifications. Both before and after a 32 hour vibration fatigue test (approx. 9 200 000 cycles of vibration) we did the following measurements on each sample.
- Output signal measurement at 5.4% excitation
- Current consumption in operation
- Sensing element resonance frequency measurement

Although the number of samples available was relatively small, the results were evaluated statistically.

*4.2.1 Output signal measurement at 5.4% excitation*

In this measurement the operability of the sample is checked and the magnitude of the output signal at 5.4% (of the 20g) excitation is recorded. The excitation is very small in this case, it equals approximately 1.37g. This way we could get an undistorted output signal, which hopefully represents the operability of the device. Fig 5, 6 and 7 show a comparison of the output signals for both before and after tests.

The difference between mean values was less than 0.007 V in every direction. This is considered as a measurement error. This way, no change was experienced between the two states. The statistical dispersion of the output signal values was less than 3% for both before and after testing.

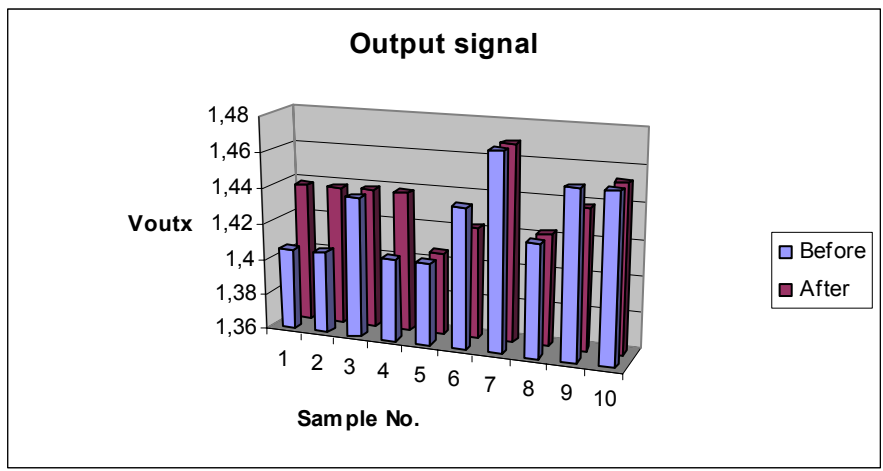

Fig. 5. Output signal Voutx before and after test

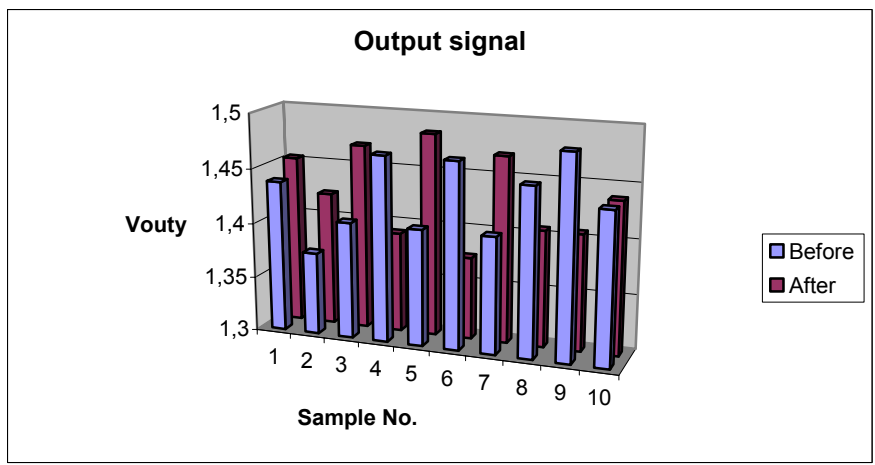

Fig. 6. Output signal Vouty before and after test

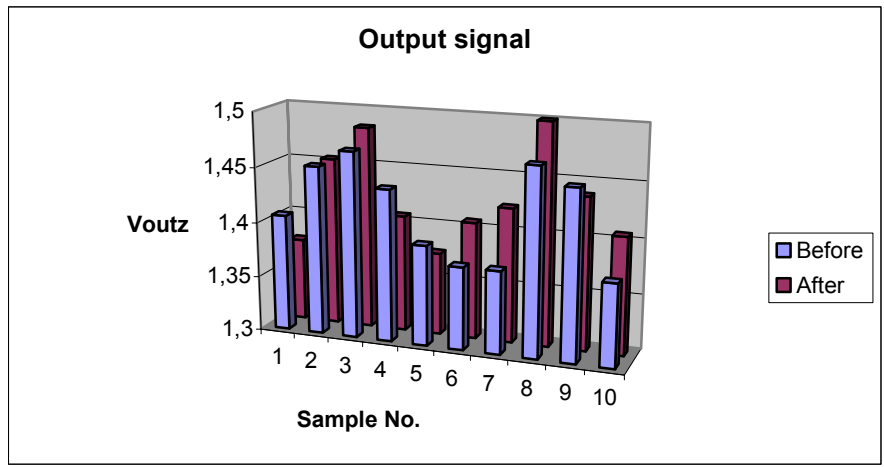

Fig. 7. Output signal Voutz before and after test

*4.2.2 Resonance frequency measurements*

A considerable resonance frequency shift, caused by vibration fatigue, could reveal material parameter instabilities. In this way, resonance frequency measurements were also carried out both before and after vibration tests, using a digital oscilloscope and the sweep mode of the function generator. As the amplification at resonance frequency can be very high compared to other values at different frequencies, only 0.78% of the total 20g acceleration was applied as an excitation signal. Fig 8, 9 and 10 show a comparison for the resonance frequency values of the different samples in directions X, Y and Z respectively.

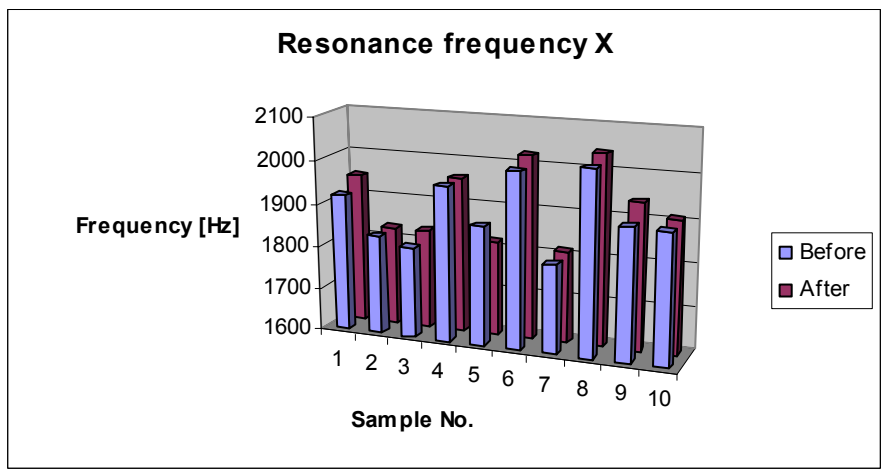

Fig. 8. Resonance frequency values for direction X

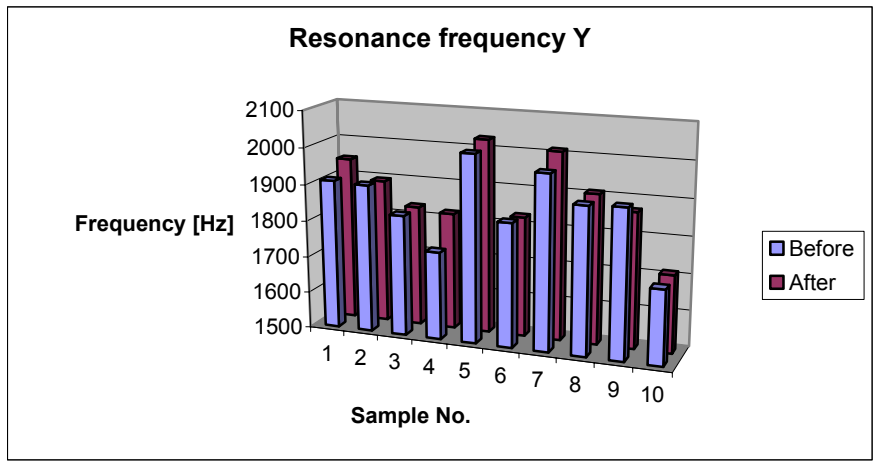

Fig. 9. Resonance frequency values for direction Z

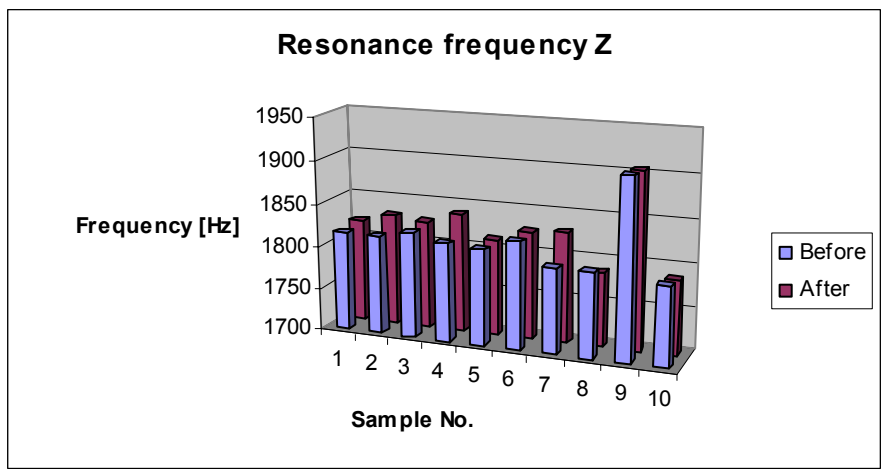

Fig. 10. Resonance frequency values for direction Z

The highest statistical dispersion of the resonance frequency between the different entities was 5.14% in direction X and the lowest 1.8% in Z. It was observable that the distribution of resonance frequency values in direction Z is almost 3 times more homogenous than for X and Y. This can be due to technological reasons. A possible explanation for this may be that the sensing properties of the mechanical structure in direction Z are related to the with of the sensing layer (which can be very homogenous within a wafer or even in a whole technology), while in the other directions these properties supposed to be defined by other circumstances like for e.g. precise mask alignment or underetch phenomena.

### 5. CONCLUSIONS

A small size, simple and cheap vibration fatigue test equipment was designed and realized at BUTE and vibration fatigue tests were carried out on 10 samples of a LIS0L02AS4-type MEMS 3-axis inertial sensor provided by ST Microelectronics. The results showed that the arrangement is easily applicable for MEMS vibration fatigue testing. No significant change in the electrical characteristics of the samples was observed and no considerable resonance frequency shift occurred due to 9 200 000 cycles of vibration.

In the final version of our paper the detailed results of the above mentioned tests will be presented. A final comparison with the results of other teams can hopefully be made at the end.

### 6. ACKNOWLEDGMENTS

This work was supported by the PATENT IST-2002-507255 Project of the EU and by the OTKA-TS049893 project of the Hungarian Government.